# High Performance 4H-SiC Optically Controlled MOS Transistor

Sitian Chen, Ziqian Tian, Guoliang Zhang, Jiafa Cai, Rongdun Hong, Xiaping Chen, Dingqu Lin, Shaoxiong Wu, Yuning Zhang, Feng Zhang

*Abstract*— This paper introduces an optically controlled 4H-SiC MOSFET designed to avoid the gate-oxide interface unreliability and electromagnetic interference (EMI) susceptibility inherent in conventional voltage-driven devices. By replacing the conventional gate electrode with a semi-transparent optical window, the device enables direct modulation of channel conductivity through ultraviolet illumination. Electrical and optical characterization demonstrates that under an optical power density above 0.1 W/cm$^2$, the device achieves an on/off current ratio exceeding 10$^6$ between illuminated and dark states. Notably, at an optical power density of 0.031 W/cm$^2$, the photogenerated current density exceeds that obtained under a gate bias of 15 V in magnitude. Energy band analysis confirms that the optical switching mechanism operates through direct photogenerated carrier generation and transport, fundamentally differing from conventional gate voltage control and thus circumventing interface-trap and EMI-related limitations. Dynamic measurements further reveal fast switching capability, with a rise time of 1.44 ns. These results validate the feasibility of optically driven switching in SiC-based devices and highlight their potential for high-speed logic applications.

## I. Introduction

Recent advances in high-performance electronic systems demand greater computing speed, reliability, and noise immunity of integrated circuits. Silicon carbide (SiC), as a representative wide-bandgap semiconductor, offers a high critical breakdown field, superior thermal conductivity, and excellent thermal stability[1], [2], [3], [4]. These properties make SiC-based MOSFETs attractive for high-frequency microelectronics and radiation-hardened switching circuits[5], [6], where they outperform conventional silicon devices[7], [8].

This research was supported by the National Natural Science Foundation of China (Grant No. 62274137 and No.62174143), National Key Research and Development Program of China (Grant No. 2023YFB3609500, Grant No.2023YFB3609502), Jiangxi Provincial Natural Science Foundation (Grant No. 20232BAB202043), and the Fundamental Research Funds for the Central Universities (No. ZK1245).

Sitian Chen, Ziqian Tian, Guoliang Zhang, Jiafa Cai, Rongdun Hong, Xiaping Chen, Dingqu Lin, Shaoxiong Wu, Yuning Zhang and Feng Zhang are with the Department of Physics, College of Physical Science and Technology, Xiamen University, Xiamen 361005, China (E-mail: fzhang@xmu.edu.cn).

Nevertheless, conventional SiC MOSFETs that rely on gate voltage for channel modulation face inherent limitations. Their performance depends critically on gate dielectric quality. Fixed charges at the SiC/SiO$_2$ interface can induce threshold voltage drift[9], [10], leading to ambiguous logic levels and degraded accuracy. Voltage-controlled operation is also susceptible to EMI. In high-frequency environments, external noise and internal interference can compromise gate signal integrity and threaten system stability[11], [12]. Switching speed is further limited by internal capacitor charging and interface trap effects. High-density traps at the SiC/SiO$_2$ interface capture carriers during turn-on, reducing channel mobility and prolonging switching transients [13], [14]. Optical control offers a promising alternative by replacing electrical signals with light. This approach circumvents gate dielectric issues while enhancing EMI immunity and electrical isolation, improving device reliability. It also aligns with the growing trend toward optical computing. Early demonstrations in materials such as GaN further underscore its potential[15], [16].

This work focuses on optically controlled switching for logic-level applications. To this end, we design and fabricate a novel optically controlled microelectronic switch based on a lateral 4H-SiC MOSFET structure. By incorporating an ultrathin transparent gate as an optical window, the device utilizes low-power UV illumination to directly generate electron–hole pairs in the channel, enabling optical on/off switching without electrical gate signals.

## II. Device Structure and Fabrication

The device was fabricated on an N-type 4H-SiC epitaxial wafer with a 10-μm-thick N$^-$ epitaxial layer ($N_D = 5 \times 10^{15}$ cm$^{-3}$) and a 365-μm-thick N$^+$ substrate ($N_D = 5 \times 10^{18}$ cm$^{-3}$). A lateral MOSFET structure with a 120 × 35 μm$^2$ photosensitive window was implemented, as illustrated in Fig. 1(a). Aluminum ion implantation formed P-wells with a doping concentration of approximately $1 \times 10^{19}$ cm$^{-3}$ and a junction depth of 0.35 μm. An interdigitated geometry with ~2.6 μm finger spacing was designed to enlarge the photosensitive area and improve carrier collection efficiency. Subsequently, a 40-nm-thick gate oxide was grown by thermal oxidation. Ohmic contacts were formed by depositing a Ti(20 nm)/Al(40 nm)/Ti(10 nm)/Au(100 nm) multilayer stack via magnetron sputtering, followed by rapid thermal annealing in argon. The semi-transparent electrode was fabricated as a 10-nm Ti/Au bilayer. Finally, Ti(10 nm)/Au(120 nm) was sputtered for bonding pads.

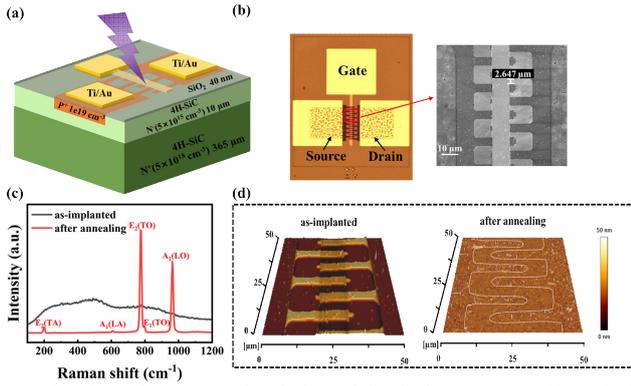

Fig. 1. (a) Schematic cross-sectional view of the device structure. (b) Optical microscope image of the device. (c) Raman spectra of the source–drain channel region before and after high-temperature activation following ion implantation. (d) AFM images of the source–drain channel region before and after high temperature activation following ion implantation.

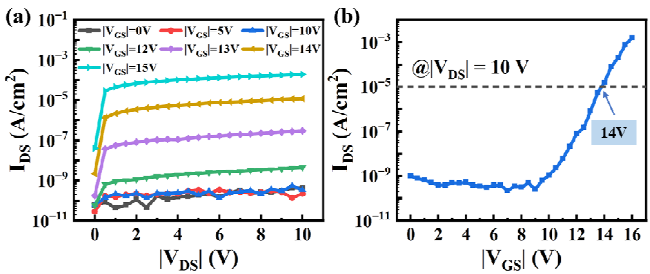

Fig. 2. Electrical measurements of the fabricated device. (a) Output characteristics with various $V_{GS}$. (b) Transfer characteristics under dark state and light at $|V_{DS}| = 10$ V.

To assess implantation damage and recovery, atomic force microscopy (AFM) and Raman spectroscopy were employed. AFM revealed ~24 nm surface protrusions in implanted regions (Fig. 1(d)), attributed to lattice amorphization and volumetric expansion resulting from high-dose ion bombardment. After annealing, these features largely disappeared, indicating effective recrystallization and planarization[17]. Raman spectra (Fig. 1(c)) showed broad amorphous SiC peaks in as-implanted samples, while annealed samples exhibited restored crystalline 4H-SiC peaks, confirming successful lattice repair[18].

Electrical and optical measurements were performed at room temperature using a Keithley 4200-SCS parameter analyzer. A xenon lamp with monochromator provided specific wavelengths, and a 266 nm picosecond pulsed laser (10 ps pulse width, 1 MHz repetition rate) with a Tektronix MSO44 oscilloscope (1 GHz bandwidth, 6.25 GS/s sampling rate) was used for dynamic response evaluation.

## III. RESULTS AND DISCUSSION

Fig. 2 shows the dark-state electrical characteristics. For gate voltages $|V_{GS}| < 10$ V, the device remains in a stable off-state, with a drain current density $1\sim 10^{-10}$ A/cm², which corresponds to a defined logic "low" level and helps minimize static power dissipation. Defining the logic "high" threshold at 10 μA/cm² requires $|V_{GS}|\approx 14$ V at $|V_{DS}| = 10$ V. At $|V_{GS}| = 15$ V, the on-state current reaches ~$2\times10^{-4}$ A/cm², yielding an on/off current ratio of up to $10^6$. This high ratio ensures robust noise margins for reliable logic operation. The extracted threshold voltage $|V_{th}|$ (at 10 μA/cm²) is approximately 14 V.

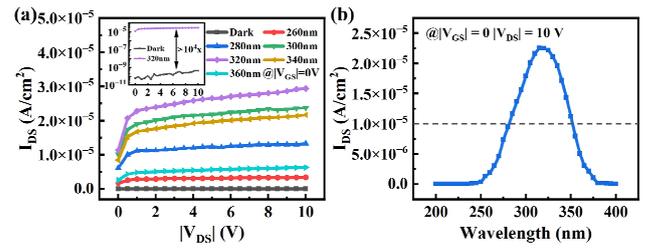

Fig. 3. Typical photoelectric properties of the 4H-SiC optically controlled MOSFET. (a) The output curves under different optical wavelengths. (b) The transfer curve as a function of wavelength at $|V_{DS}| = 10$ V.

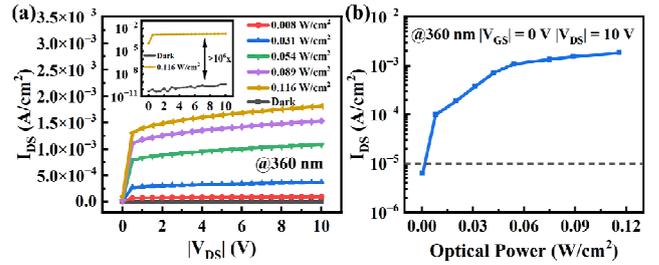

Fig. 4. Characteristics of the optically controlled 4H-SiC MOSFET under 360 nm illumination at various optical powers: (a) Output curve; (b) Transfer curve at $|V_{DS}| = 10$ V.

As shown in Fig. 3, on-state current depends strongly on wavelength, enabling wavelength-encoded logic. Under 320 nm illumination at 0.36 mW/cm², photocurrent density increases by over $10^4$ compared to the dark state, consistently exceeding the logic high threshold of 10 μA/cm². Although this value is lower than that achieved under a 15 V gate bias, this is due to the low optical power used. Comparison of Figs. 2 and 3 shows that in the 260–360 nm range, even with the xenon lamp delivering only tens of nano-watts, the resulting current density is comparable to that under ~14 V gate bias, demonstrating low-power driving potential.

Spectral response is governed by 4H-SiC absorption and device design. The device responds most strongly in the 280–360 nm range. At short wavelengths (200–280 nm), high absorption confines carrier generation near the surface, where rapid recombination at defect sites limits the number of carriers reaching the channel[19]. At longer wavelengths (360–400 nm), absorption depth increases, shifting generation below the channel, where weaker built-in field and longer transport paths reduce carrier separation efficiency. Beyond ~380 nm, photon energy falls below the 4H-SiC bandgap (~3.26 eV), causing a sharp response drop. Thus, optimal performance occurs in the 280–360 nm range, where photons penetrate the surface recombination zone while being absorbed in the channel depletion region, enabling efficient carrier generation, separation, and collection.

Fig. 4(a) displays the I–V characteristics under 360 nm illumination at various power densities. Increasing optical power raises photon flux, exciting more carriers and increasing photocurrent. Notably, at a power density of 0.031 W/cm², the optically controlled current density ($3.7\times10^{-4}$ A/cm²) surpasses that achieved under $|V_{GS}| = 15$ V, underscoring optical driving potential. At powers above 0.1 W/cm², the device achieves an on/off current ratio exceeding $10^6$, which sharpens the logic transition and enhances noise immunity in complex circuits.

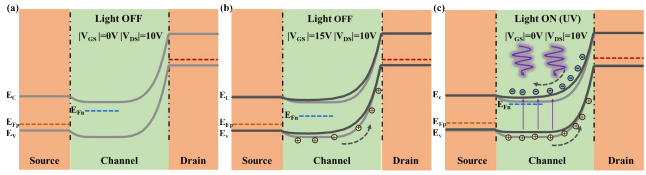

Fig. 5. Schematic band diagrams of the device at $|V_{DS}| = 10$ V (a) Under dark conditions. (b) With applied gate bias. (c) Under ultraviolet light illumination.

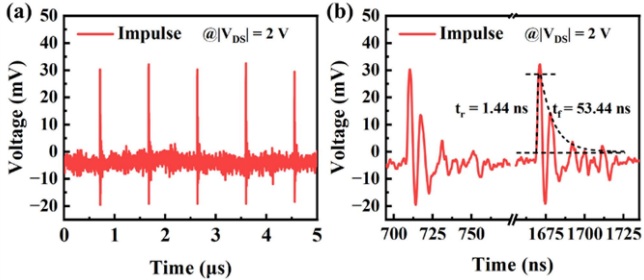

Fig. 6. Transient response curve of the optically controlled 4H-SiC MOSFET at $|V_{DS}| = 2$ V: (a) multiple-pulse and (b) double-pulse.

The operating principles under electrical and optical control are illustrated in the energy band diagram of Fig. 5. At the off-state, a reverse-biased PN junction between P⁺ regions and the N⁻ epitaxial layer establishes a high potential barrier, resulting in high channel resistance. Under electrical control, sufficient gate voltage ($|V_{GS}| > |V_{th}|$) bends the bands upward, lowering the potential barrier and enabling efficient hole transport along the lateral electric field, thereby turning the device on. This mechanism depends critically on gate dielectric quality and the integrity of the gate-induced electric field. In contrast, under optical control, ultraviolet photons directly generate a high density of electron-hole pairs within the channel and depletion region, rapidly increasing the mobile carrier concentration and reducing channel resistance. Accumulation and extraction of these photogenerated carriers partially screen the built-in electric field and modify the band profile, lowering the energy barrier. This direct conductivity modulation bypasses gate-induced inversion, inherently mitigating interface trap and EMI issues. Energy band analysis confirms fundamentally distinct turn-on mechanisms, with optical switching operating via direct photogeneration and electrical switching relying on gate-field-induced carrier inversion.

Switching speed is critical for high-frequency applications. As shown in Fig. 6, the device exhibits a rise time of 1.44 ns and a fall time of 53.44 ns under 266 nm illumination. The nanosecond-scale response time demonstrated here is consistent with other reports of high-speed SiC optoelectronic devices[20]. This asymmetry arises from distinct turn-on and turn-off mechanisms. The fast rise time results from rapid photogeneration of carriers and their subsequent drift in external electric field. This optical turn-on mechanism inherently bypasses gate-capacitance charging and interface trap filling that limit conventional MOSFETs, allowing turn-on time (<2 ns) to surpass that of conventional SiC MOSFETs (tens to hundreds of nanoseconds[21], [22]). Moreover, since photogenerated carriers are created throughout the channel depletion region rather than being confined to the 4H-SiC/SiO₂ interface, they experience reduced scattering and trapping at interface states, further accelerating the turn-on transient. The slower fall time is primarily dictated by carrier recombination lifetime. After illumination ceases, excess carriers dissipate through relatively slow recombination channels such as Shockley–Read–Hall(SRH) and surface recombination[23], [24]. Parasitic capacitance in the measurement setup also contributes to the extended fall transient. Finally, Fig. 7 benchmarks our proposed device against other optically controlled devices[16], [25], [26], [27], [28], [29], highlighting its superior combination of low optical power and fast switching speed.

| Device Type | Optical power (mW) | $V_{DS}$ (V) | $V_{GS}$ (V) | $I_{DS}$/Optical power (A/W) | $t_{total}$ (ns) | Ref |
|---|---|---|---|---|---|---|
| SiC MOSFET | **0.0049** | 10 | 0 | $1.55\times10^{-2}$ | **54.88** | This work |
| SiC MOSFET | 44 | 1500 | 18.7 | 68 | 367 | [25] |
| SiC MOSFET | 4000 | 500 | 19 | 0.1925 | 300 | [26] |
| GaN FinFET | 0.0106 | 1 | 0 | 589.6 | >10⁶ | [27] |
| GaN HEMT | 245 | 50 | 0 | $4.49\times10^{-2}$ | 2473 | [16] |
| GaN HEMT | 90 | 400 | 4.4 | 33 | 127.2 | [28] |
| Si MOSFET | 4000 | 50 | 12 | 0.3125 | 310 | [29] |

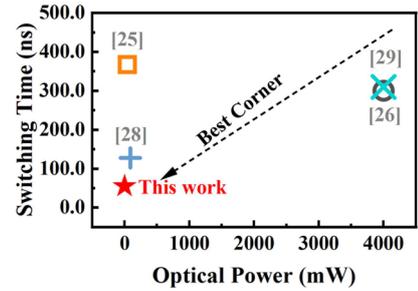

Fig. 7. (Top) Comparison of key performance metrics between this work and previously reported optically controlled devices. (Bottom) Benchmark of total switching time ($t_r+t_f$) versus optical power.

## IV. Conclusion

In this work, a lateral 4H-SiC MOSFET with a semi-transparent optical gate was designed and fabricated for photonic switching. Under 360 nm ultraviolet illumination at 0.031 W/cm², the device delivers a current density of ~3.7×10⁻⁴ A/cm², surpassing the output achieved under 15 V gate bias. At optical power densities above 0.1 W/cm², the device attains an on/off current ratio exceeding 10⁶ between illuminated and dark states. Dynamic characterization shows nanosecond switching with 1.44 ns rise and 53.44 ns fall times. These results demonstrate the promise of optically driven SiC devices for high-speed, EMI-immune operation in future electronic systems.

## References

[1] P. Sharmila, G. Supraja, D. Haripriya, C. Sivamani, and A. L. Narayana, "Silicon carbide MOSFETs: A critical review of applications, technological advancements, and future perspectives," *Micro and Nanostructures,* vol. 202, Jun 2025, Art no. 208126, doi: 10.1016/j.micrna.2025.208126.

[2] J. Liu, "From Silicon to Wide Bandgap Semiconductors: Exploration of Compound Semiconductors in High Frequency and High Power Devices," *Applied and Computational Engineering,* vol. 162, no. 1, pp. 209-214, Jun 2025, doi: 10.54254/2755-2721/2025.GL24406.

[3] C. Yang, S. S. Wei, and D. J. Wang, "Bias temperature instability in SiC metal oxide semiconductor devices," *Journal of Physics D-Applied Physics,* vol. 54, no. 12, Mar 2021, Art no. 123002, doi: 10.1088/1361-6463/abcd5e.

[4] X. G. Zhu, Z. W. Shen, Z. J. Wang, Z. R. Liu, Y. Y. Miao, S. Z. Yue, Z. Fu, Z. H. Li, Y. N. Zhang, R. D. Hong, S. X. Wu, X. P. Chen, J. F. Cai, D. Y. Fu, and F. Zhang, "Effects of gamma-ray irradiation on material and electrical properties of AlN gate dielectric on 4H-SiC," *Nanotechnology,* vol. 35, no. 27, Jul 2024, Art no. 275704, doi: 10.1088/1361-6528/ad3a6e.

[5] F. Roccaforte, P. Fiorenza, G. Greco, R. Lo Nigro, F. Giannazzo, F. Iucolano, and M. Saggio, "Emerging trends in wide band gap semiconductors (SiC and GaN) technology for power devices," *Microelectronic Engineering,* vol. 187-188, pp. 66-77, Feb 2018, doi: 10.1016/j.mee.2017.11.021.

[6] X. Li, X. Li, P. K. Liu, S. X. Guo, L. Q. Zhang, A. Q. Huang, X. C. Deng, and B. Zhang, "Achieving Zero Switching Loss in Silicon Carbide MOSFET," *IEEE Transactions on Power Electronics,* vol. 34, no. 12, pp. 12193-12199, Dec 2019, doi: 10.1109/TPEL.2019.2906352.

[7] C. Peng, H. Zhang, Z. G. Zhang, T. Ma, Z. Z. Wang, and Z. F. Lei, "Comparison of 14-MeV Neutron-Induced Damage in Si and SiC Power MOSFETs," *IEEE Transactions on Electron Devices,* vol. 72, no. 9, pp. 5104-5110, Sep 2025, doi: 10.1109/TED.2025.3588129.

[8] H. Z. Huang, N. Y. Wang, J. L. Wu, and T. B. Lu, "Radiated disturbance characteristics of SiC MOSFET module," *Journal of Power Electronics,* vol. 21, no. 2, pp. 494-504, Feb 2021, doi: 10.1007/s43236-020-00187-4.

[9] V. Volosov, S. Cascino, M. Saggio, A. Imbruglia, F. Di Giovanni, C. Fiegna, E. Sangiorgi, and A. N. Tallarico, "Role of interface/border traps on the threshold voltage instability of SiC power transistors," *Solid-State Electronics,* vol. 207, Sep 2023, Art no. 108699, doi: 10.1016/j.sse.2023.108699.

[10] T. Kimoto and H. Watanabe, "Defect engineering in SiC technology for high-voltage power devices," *Applied Physics Express,* vol. 13, no. 12, Dec 2020, Art no. 120101, doi: 10.35848/1882-0786/abc787.

[11] A. K. Morya, M. C. Gardner, B. Anvari, L. M. Liu, A. G. Yepes, J. Doval-Gandoy, and H. A. Toliyat, "Wide Bandgap Devices in AC Electric Drives: Opportunities and Challenges," *IEEE Transactions on Transportation Electrification,* vol. 5, no. 1, pp. 3-20, Mar 2019, doi: 10.1109/TTE.2019.2892807.

[12] Y. Z. Wu, S. Yin, H. Li, and M. H. Dong, "Modeling and Experimental Investigation of Electromagnetic Interference (EMI) for SiC-Based Motor Drive," *Energies,* vol. 13, no. 19, Oct 2020, Art no. 5173, doi: 10.3390/en13195173.

[13] M. Chaturvedi, S. Dimitrijev, H. A. Moghadam, D. Haasmann, P. Pande, and U. Jadli, "Fast Near-Interface Traps in 4H-SiC MOS Capacitors Measured by an Integrated-Charge Method," *IEEE Access,* vol. 9, pp. 109745-109753, Aug 2021, doi: 10.1109/ACCESS.2021.3102614.

[14] K. H. Yu, Y. F. Liu, W. H. Zhang, H. A. Chen, C. Z. Li, J. Q. Ding, J. Wang, D. Y. Zhai, and Y. W. Wang, "Characterization of the slow-state traps in 4H-SiC P-type MOS capacitor by a preconditioning technique with high positive voltage stress," *Micro and Nanostructures,* vol. 175, Mar 2023, Art no. 207506, doi: 10.1016/j.micrna.2023.207506.

[15] J. H. Hsia, J. A. Perozek, and T. Palacios, "First Demonstration of Optically-Controlled Vertical GaN finFET for Power Applications," *IEEE Electron Device Letters,* vol. 45, no. 5, pp. 774-777, May 2024, doi: 10.1109/LED.2024.3375856.

[16] E. Palmese, H. T. Xue, D. J. Rogers, and J. J. Wierer, Jr., "Light-Triggered, Enhancement-Mode AlInN/GaN HEMTs With Sub-Microsecond Switching Times," *IEEE Electron Device Letters,* vol. 45, no. 10, pp. 1903-1906, Oct 2024, doi: 10.1109/LED.2024.3440177.

[17] V. Boldrini, A. Parisini, and M. Pieruccini, "Analysis of the electrical activation data in thermally annealed implanted Al/4H-SiC systems: A novel approach based on cooperativity," *Materials Science in Semiconductor Processing,* vol. 148, Sep 2022, Art no. 106825, doi: 10.1016/j.mssp.2022.106825.

[18] J. O. Orwa, K. W. Nugent, D. N. Jamieson, and S. Prawer, "Raman investigation of damage caused by deep ion implantation in diamond," *Physical Review B,* vol. 62, no. 9, pp. 5461-5472, Sep 2000, doi: 10.1103/PhysRevB.62.5461.

[19] C. V. Prasad, P. Dharmaiah, G. H. Lee, S. R. Park, J. H. Kim, S. H. Park, H. D. Kang, M. Kim, J. W. Choi, Y. J. Kim, C. J. Park, G. Tse, H. J. Park, W. H. Shin, N. K. Jaiswal, Y. K. Mishra, T. Ebel, J. M. Oh, and S. M. Koo, "Interface engineering of 4H-SiC-based UV photodetectors: A comprehensive review," *Materials Today Advances,* vol. 28, Dec 2025, Art no. 100662, doi: 10.1016/j.mtadv.2025.100662.

[20] Z. Fu, J. Liu, M. Yuan, J. F. Cai, R. D. Hong, X. P. Chen, D. Q. Lin, S. X. Wu, Y. N. Zhang, Z. Y. Wu, Z. W. Shen, Z. J. Wang, J. C. Wang, M. K. Zhang, Z. L. Yang, D. Y. Fu, F. Zhang, and R. Zhang, "Local avalanche photodetectors driven by lightning-rod effect and surface plasmon excitations," *Nature Communications,* vol. 17, no. 1, Dec 2025, Art no. 76, doi: 10.1038/s41467-025-66790-w.

[21] H. Y. Yu, J. Wang, J. Y. Zhang, S. W. Liang, and Z. J. Shen, "Theoretical Analysis and Experimental Characterization of 1.2-kV 4H-SiC Planar Split-Gate MOSFET With Source Field Plate," *IEEE Transactions on Electron Devices,* vol. 71, no. 3, pp. 1508-1512, Mar 2024, doi: 10.1109/TED.2023.3336644.

[22] D. Chae, W. Jung, K. Kim, W. Seok, and M. Shim, "A Thyristor-Structured-Based Low-Power Demodulator Circuit for High Reliability and Short-Circuit Current Reduction," *IEEE Transactions on Power Electronics,* vol. 40, no. 4, pp. 4740-4746, Apr 2025, doi: 10.1109/TPEL.2024.3520548.

[23] D. W. deQuilettes, K. Frohna, D. Emin, T. Kirchartz, V. Bulovic, D. S. Ginger, and S. D. Stranks, "Charge-Carrier Recombination in Halide Perovskites," *Chemical Reviews,* vol. 119, no. 20, pp. 11007-11019, Oct 2019, doi: 10.1021/acs.chemrev.9b00169.

[24] X. H. Hou, X. L. Zhao, Y. Zhang, Z. F. Zhang, Y. Liu, Y. Qin, P. J. Tan, C. Chen, S. J. Yu, M. F. Ding, G. W. Xu, Q. Hu, and S. B. Long, "High-Performance Harsh-Environment-Resistant GaOX Solar-Blind Photodetectors via Defect and Doping Engineering," *Advanced Materials,* vol. 34, no. 1, Jan 2022, Art no. 2106923, doi: 10.1002/adma.202106923.

[25] X. Yang, G. Shi, L. Jin, Y. Qin, M. Porter, X. Jia, D. Dong, L. Shao, and Y. Zhang, "Ultrafast Optically Controlled Power Switch: A General Design and Demonstration With 3.3 kV SiC MOSFET," *IEEE Transactions on Electron Devices,* vol. 71, no. 12, pp. 8025-8030, Dec 2024, doi: 10.1109/TED.2024.3485018.

[26] A. Meyer, S. K. Mazumder, and H. Riazmontazer, "Optical Control of 1200-V and 20-A SiC MOSFET," in *27th Annual IEEE Applied Power Electronics Conference and Exposition (APEC)*, Orlando, FL, USA, Feb 2012, pp. 2530-2533, doi: 10.1109/APEC.2012.6166179.

[27] J. H. Hsia, J. A. Perozek, J. Park, and T. Palacios, "First Demonstration of a Fully -Vertical GaN Power finFET with Direct Optical Triggering," in *37th International Symposium on Power Semiconductor Devices and Integrated Circuits-ISPSD-Annual*, Kumamoto-shi, JAPAN, Jun 2025, pp. 53-56, doi: 10.23919/ISPSD62843.2025.11118256.

[28] X. Yang, L. Y. Jin, M. Porter, H. C. Cui, Z. N. Yang, H. H. Gong, H. Wang, L. B. Shao, and Y. H. Zhang, "First Demonstration of Optically -Controlled 650 V Power GaN HEMT with Ultrafast Switching Speed," in *37th International Symposium on Power Semiconductor Devices and Integrated Circuits-ISPSD-Annual*, Kumamoto-shi, JAPAN, Jun 2025, pp. 49-52, doi: 10.23919/ISPSD62843.2025.11117348.

[29] S. K. Mazumder and T. Sarkar, "Optically Activated Gate Control for Power Electronics," *IEEE Transactions on Power Electronics,* vol. 26, no. 10, pp. 2863-2886, Oct 2011, doi: 10.1109/TPEL.2009.2034856.